\documentclass[aps,pre,twocolumn,showpacs,preprintnumbers,superscriptaddress]
{revtex4}

\usepackage{graphicx}
\usepackage{epsf} 
\usepackage{dcolumn}
\usepackage{bm}
\usepackage{amssymb}
\usepackage{amsmath}
\usepackage{amsbsy}
\usepackage{times}

\begin{document}

\title{Tumor growth instability and the onset of invasion}

\author{Mario Castro}
\affiliation{Grupo Interdisciplinar de Sistemas Complejos (GISC)
and Grupo de Din\'amica No Lineal (DNL), Escuela T\'ecnica Superior
de Ingenier{\'\i}a (ICAI), Universidad Pontificia Comillas,
E-28015 Madrid, Spain}

\author{Carmen Molina-Par\'{\i}s}
\affiliation{Department of Applied Mathematics, University of Leeds,
Leeds LS2 9JT, UK and \\
Departamento de Matem\'aticas, 
F\'{\i}sica Aplicada y
F\'{\i}sico-qu\'{\i}mica,
Facultad de Farmacia, Universidad San Pablo CEU, E-28660 Madrid, Spain}
\author{Thomas S. Deisboeck}
\affiliation{
Complex Biosystems Modeling Laboratory, Harvard-MIT (HST) 
Athinoula A. Martinos Center for Biomedical Imaging, 
Massachusetts General Hospital, Charlestown, MA 02129, USA}

\begin{abstract}

Motivated by experimental observations, we develop a mathematical
model of chemotactically directed tumor growth. We present an
analytical study of the model as well as a numerical one.  The
mathematical analysis shows that: (i) tumor cell proliferation by
itself cannot generate the invasive branching behaviour observed
experimentally, (ii) heterotype chemotaxis provides an instability
mechanism that leads to the onset of tumor invasion and (iii) homotype
chemotaxis does not provide such an instability mechanism but enhances
the mean speed of the tumor surface. The numerical results not only
support the assumptions needed to perform the mathematical analysis
but they also provide evidence of (i), (ii) and (iii). Finally, both
the analytical study and the numerical work agree with the
experimental phenomena.

\end{abstract}



\maketitle
 
\section{Introduction}
\label{sec:intro}

Experiments have shown that a variety of tumor cells produce both
protein growth factors and their corresponding receptors, enabling a
mechanism termed {\em autocrine} or {\em paracrine}, if the stimulus
not only affects the {\em source} but also its bystander
cells~\cite{ref1}.  Such polypeptide growth factors can, for example,
stimulate tumor cell growth and invasion, such as in the case of
hepatocyte growth factors~\cite{ref2} and epidermal growth
factor~\cite{ref3}, or induce tumor angiogenesis (through secretion of
vascular endothelial growth factor, VEGF~\cite{ref4}). The biological
evidence supporting these paracrine/autocrine loops suggests that such
signaling factors have significance for cell-cell
interaction~\cite{ref5}.  Depending on the cancer type, its
characteristic features include a combination of rapid volumetric
growth and genetic/epigenetic heterogeneity, as well as extensive
tissue invasion with both local and distant dissemination. Such tumor
cell motility has been intensely investigated and found to be guided
by diffusive chemical gradients, a process called chemotaxis, {\it
e.g.,}~\cite{ref6}. Since cell signaling and information processing
on the microscopic scale should also determine the emergence of both
multicellular patterns and macroscopic disease dynamics, it is
intriguing to characterize the relationship between environmental
stimuli and the cell-signaling {\em code} they trigger.

In a recent paper, Sander and Deisboeck~\cite{sander} showed that a
combined heterotype and homotype diffusive chemical signal can yield
invasive cell branching patterns seen in microscopic brain tumor
experiments~\cite{prolif} by means of a discrete model, for some
specific form of the interactions. There is a long history on the
study of this type of assay~\cite{sage-vernon,ref4}. There are also
quite a few mathematical models that address the issue of tumor growth
and cell migration~\cite{byrne-chap,murray-lubkin,preziosi,
greenspan,ward-king,holmes,maini,libro-sleeman}. In Ref.~\cite{sander}
the authors carry out a linear stability analysis from the steady
state and show that both homo and heterotype chemotaxis are required
for the development of invasive branching behaviour.  Employing an
improved version of our previously developed reaction-diffusion
model~\cite{physicaa}, we now specifically investigate the
relationship between an extrinsic nutrient signal, heterotype
chemoattractant $Q$, and the homotype soluble signal $C$, produced by
the tumor cells themselves.  The underlying oncology concept is that
in the process of spatio-temporal tumor expansion, $C$ functions as a
guiding {\em cue} for mobile cancer cells, directing the trailing ones
towards sites of higher $Q$ concentration and thus avoiding tissue
areas with low or decreasing density of $Q$. In a sense, the
dynamically changing $C$ profile, secreted by the tumor cells, encodes
the underlying $Q$ {\em map}, which in itself represents a particular
tissue environment. However, this picture is difficult to
quantitatively assess with conventional experimental assays, and it
has not yet been theoretically demonstrated in a clear-cut way.

In this paper, we are therefore particularly interested in how these
mechanisms, homo and heterotype chemotaxis, compete and/or cooperate
in the formation of tumor branching structures and how the
mathematical reaction terms must be chosen in order to reproduce, with
some degree of universality, the experimentally observed
patterns. Thus, we will study the impact of a fixed extrinsic nutrient
source, and how the interplay among nutritive, mechanical and chemical
properties support the principle of {\em least resistance, most
attraction} for spatio-temporal tumor cell
expansion~\cite{prolif}. The analytical and numerical results provide
insight as to the cells' ability to readily modulate $C$, as well as,
more generally, to the importance of paracrine growth factors for
information transfer in multicellular biosystems. We have kept the
mathematical model as general as possible in order to understand the
essential features of the time evolution of the system.

The report of our results is organized as follows. We describe our
reaction-diffusion mathematical model in Sec.~\ref{sec:model}, where a
detailed discussion of each of the involved mechanisms is given
separately. Section~\ref{sec:analy} reports general analytical results
for the model of Sec.~\ref{sec:model}.  We numerically check the
validity of our analytical results in Sec.~\ref{sec:num}. Finally, we
conclude in Sec.~\ref{sec:end} with a discussion of our results and
provide a picture of chemotactic cell invasion as triggered by tumor
growth instability.

\section{Mathematical model}
\label{sec:model}

Tumor expansion is a multi-step process that involves, in a
non-trivial fashion, several mechanisms of progression. Here, we
concentrate on tumor cell proliferation and chemotactically guided
invasion, induced by both a diffusive heterotype and a homotype
attractant (produced by the very tumor cells).

\subsection{Extracellular matrix gel (matrigel)}
\label{sec:M}

The experimental setting, which we have modeled here, consisted of a
multicellular tumor spheroid (MTS) embedded in a tissue culture
medium-enriched extracellular matrix (ECM) gel,
Matrigel~\cite{prolif}. We consider $M=M({\bf x},t)$, the average
density field of the gel matrix as a function of space ${\bf x}$ and
time $t$. The role of $M$ is twofold. From a nutrient perspective, the
tumor cells (whose concentration will be described by the density
field $U({\bf x},t)$ hereafter) metabolize $M$ and hence they are able
to proliferate. Besides, from a mechanical perspective, the solid gel
matrix has an impact on tumor cell mobility, {\em i.e.,} it confines
the tumor cells and so they are guided by {\em least or lesser
resistance} areas throughout the ECM here in vitro, or, in vivo, by
the distinct mechanical properties of the surrounding
tissue. Therefore, at a particular site, more $M$ can sustain a higher
concentration of tumor cells, which in turn will metabolize more of
the nourishing gel medium and thus, over time, will lead to an on-site
reduction of the ECM matrix' mechanical resistance, which had
initially hindered cell motility.

Mathematically, we assume that the consumption rate of $M$ by $U$,
which we denote by $R_M(M,U)$, grows monotonically with both
variables, and is a non-negative function. Later, we will make some
assumptions regarding the mechanical impact of $M$ on the diffusivity
of tumor cells and of both heterotype and homotype chemoattractants.

For the sake of generality, we assume that the matrigel medium can
diffuse, with constant diffusivity $\mu_M$. The order of magnitude of
$\mu_M$ depends on the specific type of medium under consideration
(see Sec.~\ref{subsec:prolif} for further discussion on this issue).
In summary, we can write the following equation for the matrigel $M$
\begin{equation}
\partial_t M=\mu_M \nabla^2 M
-\lambda_M R_M(M,U),
\label{basic-m}
\end{equation}
where $\lambda_M$ is the inverse of the characteristic time of the $M$
consumption process. Equation~(\ref{basic-m}) reflects the fact that
the matrigel nutrient is metabolized by the tumor cells and not
replenished.

\subsection{Heterotype chemoattractant}
\label{sec:Q}

Chemotaxis can be generally defined as motility induced and guided by
a concentration gradient. As in our previous model~\cite{physicaa},
the heterotype chemoattractant represents nutrients diffusing from a
source, {\em e.g.,} in vivo, a blood vessel, and as such is what
should guide both on-site cell proliferation and the onset of
invasion.  Chemotaxis has been extensively studied in the
literature~\cite{chemotaxis1,chemotaxis2}.  It is generically assumed
that the chemotactic flux takes the form
\begin{displaymath}
{\bf J}_{chem}=\chi_Q(Q,M)U\nabla Q.
\end{displaymath}
Note that this flux is proportional to the tumor cell concentration
$U$. The function $\chi_Q(Q,M)$ is usually called the chemotactic
sensitivity, and is a positive decreasing (or at least constant)
function of both arguments $Q$ and $M$.  The explicit dependence of
$\chi_Q$ on $M$ reflects the effect of the mechanical {\em pressure}
of the underlying medium, that constrains both tumor cell and
heterotype chemoattractant movement. Besides, tumor cells digest
chemoattractant molecules, so as the former move towards a positive
gradient of $Q$, the concentration of $Q$ diminishes. This reduction
is governed by the reaction term $R_Q(Q,U)$.

Combining these ideas, we obtain the following equation for the
heterotype chemoattractant field density $Q$:
\begin{equation}
\partial_t Q= \nabla (\mu_Q(M) \nabla Q) -a_Q R_Q(Q,U),
\label{basic-q}
\end{equation}
where $a_Q$ is the inverse of the characteristic time of the $Q$
consumption process.

In the experimental in vitro setting modeled here, the heterotype
chemoattractant was supplied externally. Acknowledging that the
original experimental setting~\cite{prolif} used a non-replenished
nutrient source, here, for simplicity, we model a replenished source
of $Q$ and equation~(\ref{basic-q}) thus has to be supplemented
accordingly. This can be modeled by means of the following boundary
condition:
\begin{equation}
Q(x=L,t)=Q_0,
\label{bc1d}
\end{equation}
for one-dimensional systems, where $L$ is the size of the system, and
\begin{equation}
Q(x=L_x,y,t)=Q_0,
\label{bc2d}
\end{equation}
for two-dimensional ones, where $L_x$ is the horizontal size of the
system and $L_y$ the vertical one.

\subsection{Homotype chemoattractant}
\label{sec:C}

Tumor cells have been shown to produce protein growth factors such as
the transforming growth factor alpha (TGF-$\alpha$). These growth
factors can affect the tumor producing cell itself, hence generate an
``autocrine'' feedback loop, as well as bystander cells, an effect
called ``paracrine''~\cite{paracrine}. In the following, we refer to
this soluble chemical effector, as homotype chemoattractant and we
denote by $C$ its density field.

Since an ever growing population of tumor cells digests more $Q$, the
homotype chemoattractant $C$ may take over at some point as {\em
guidance cue} in the regions with low $Q$ concentration.  First, we
assume that the homotype chemoattractant is both released and
internalized, or (for the purposes here), taken up or consumed by the
tumor cells. The latter is based on a ligand-receptor interaction and
thus on internalization of the class of protein growth factors, which
$C$ represents. Note, that if all cells produce $C$, a cell close to
the main tumor would be less {\em inclined} to move away from it,
since it is close to a large basin of $C$. One could tune the
production rate of $C$ in such a way as to ensure that only the
density profile of $C$ near the tumor surface has an impact on the
``decision'' of a tumor cell to stay (proliferate) or to start moving
(invasion). This effect should have an impact on the tumor cell
density $U$.

The above discussion implies that $C$ is also chemotactic for $U$. The
main difference with the heterotype chemoattractant $Q$ is that the
homotype chemoattractant $C$ is produced (and consumed) by the tumor
cells. We denote by $R_C^{(p)}(M,U)$ and $R_C^{(d)}(C,U)$ the
production and digestion rates of $C$, respectively. Earlier studies
have shown~\cite{necrosis1,necrosis2,necrosis3} that eventually a
central dead area develops due to the lack of nutrients inside the
tumor spheroid, which in turn leads to the release of growth
inhibitory factors from the dying cells~\cite{necrosis4}.  A full
consideration of the development of such a {\em necrotic core} is out
of the scope of this paper~\cite{necrotic}. We consider the existence
of this ``dead area'' in the dependence of $R_C^p$ on the matrigel
density $M$. This dependence means that the production of $C$ is
enhanced where $M$ is high (outside the main tumor, as inside the
tumor the matrigel has been degraded) and therefore, the production of
$C$ is maximized for {\em reactive} tumor cells, {\em i.e.,} surface
tumor cells outside the necrotic core of the tumor~\cite{prolif}.  We
also assume that $\chi_Q(Q,M)$ is typically larger than the
chemotactic sensitivity of $C$, $\chi_C(C,M)$, for the concentrations
involved in the problem and that $\chi_C(C,M)$ depends both on $C$ and
$M$. The explicit dependence of $\chi_C$ on $M$ reflects the fact that
the matrigel constrains homotype chemoattractant movement as well.
Finally, $C$ also diffuses with diffusion coefficient $\mu_C(M)$.

In summary, the homotype chemoattractant obeys the following equation:
\begin{equation}
\partial_t C= 
\nabla(\mu_C(M) \nabla C)+\alpha_{C}R_C^{(p)}(M,U)-a_{C}R_C^{(d)}(C,U),
\label{basic-c}
\end{equation}
where $\alpha_C$ and $a_C$ are the inverse of the characteristic time
of the $C$ production and consumption process, respectively.

\subsection{Tumor cells}
\label{sec:U}

The {\em global nutrient} density available to tumor cells is
proportional to the medium density $M$. We then assume that tumor
cells proliferate with a rate that is proportional to the rate of
consumption of $M$. Moreover, we consider that tumor cells diffuse
with a diffusion constant $\mu_U$, and that $\mu_U$ depends on $M$ to
reflect the mechanical {\em pressure} of the matrigel $M$.  As {\em
tumor cells are much larger than the chemoattractant molecules} we
have $\mu_Q > \mu_C > \mu_U$.  As we discussed above, tumor cells move
towards positive gradients of both hetero and homotype
chemoattractants, so we can write
\begin{eqnarray}
\partial_tU&=& \nabla(\mu_U(M) \nabla U)-
\nabla(U \chi_Q(Q,M) \nabla Q) \nonumber \\
&&- \nabla(U \chi_C(C,M) \nabla C) +\lambda_U R_M(M,U),
\label{basic-u}
\end{eqnarray}
where $\lambda_U$ is the inverse of the characteristic time of tumor
proliferation.  Equations~(\ref{basic-m})-(\ref{basic-u}) constitute
our reaction-diffusion tumor growth mathematical model.

\section{Analytical study}
\label{sec:analy}

As we have stated above, the precise relevance of each of the factors
summarized in Sec.~\ref{sec:model} is not clearly understood. Partly,
this is due to the complex interaction amongst these factors but,
mainly, due to the lack of a systematic analytical and numerical
analysis of each individual mechanism operating in the full system.

In this section, we analyze in detail every mechanism involved in
tumor growth to determine which conditions trigger the formation of
invasive branches, and how the interplay among those mechanisms
allows these branches to be sustained in time.

\subsection{Growth due to cell proliferation}
\label{subsec:prolif}

Consider the subsystem of equations formed by Eq.~(\ref{basic-m}) and
Eq.~(\ref{basic-u}) with $\chi_Q=\chi_C=0$, defined in a d-dimensional
volume $V$.  The tumor and nutrient {\em particles} are confined into
the system and so the flux of material through the boundaries
vanishes.  This physical constraint introduces a conservation law in
the problem, namely,
\begin{equation}
\frac{d}{dt}\int_V d{\bf x}(\lambda_MU+\lambda_UM)=0,
\label{M_U_1}
\end{equation}
independently of the precise functional form of the reaction term
$R_M(M,U)$.

In the absence of diffusion ($\mu_M=\mu_U=0$) this conservation law
provides the following closed relation
\begin{equation}
M=K-\lambda_M/\lambda_UU, 
\label{M_U}
\end{equation}
where $K$ depends on the initial conditions of $M$ and $U$. If the
diffusion coefficients do not vanish, we cannot obtain such closed
relation between $U$ and $M$, except in some simpler cases (related to
the geometry of the volume and the initial conditions). With no loss
of generality, we restrict ourselves to one and two-dimensional
systems, $V=L$ and $V=L_x \times L_y$, respectively, and initial
conditions such that $M({\bf x},t=0)=0$ where $U({\bf x},t=0)\neq 0$
and $U({\bf x},t=0)=0$ where $M({\bf x},t=0)\neq0$. Then, we can
obtain a relation similar to Eq.~(\ref{M_U}). Physically, this means
that initially the nutrient surrounds the implanted tumor. In this
case, Eq.~(\ref{M_U}) is valid after a small transient time (see
Sec.~\ref{sec:num}), although the shape of the front ({\em i.e.,
surface of the tumor}) changes slightly. However, as we are interested
in the case where tumor cells diffuse slowly, this front will be
assumed to be sharp, and hence its exact shape is not relevant for our
discussion below~\cite{greenspan}. Thereby, we simply get
\begin{equation}
\partial_tU= \nabla(\mu_U( K-\lambda_M/\lambda_UU)\nabla U) 
+\lambda_U R_M(K-\lambda_M/\lambda_UU,U).
\label{prolif}
\end{equation}
Tumor cells digest the matrigel nutrient when they are in direct
contact. Thus, the reaction term $R_M(M,U)$ must vanish when any of
its arguments does. The simplest choice of such a reaction term is
$R_M(M,U)=MU$. Although other choices are possible (leading
qualitatively to the same results), we consider this choice to
illustrate the main properties of the reduced system given by
Eq.~(\ref{prolif}).

At this stage of the formal presentation, we need to consider
separately the cases $\mu_M\ll \mu_U$ and $\mu_M\gg \mu_U$. Note that
$\mu_U$ is $M$-dependent so these inequalities need to be understood
in an average sense.

\subsubsection{$\mu_M\ll \mu_U$}

\begin{figure}[!hbt] 
\begin{center}
\includegraphics[width=0.3\textwidth,clip=]{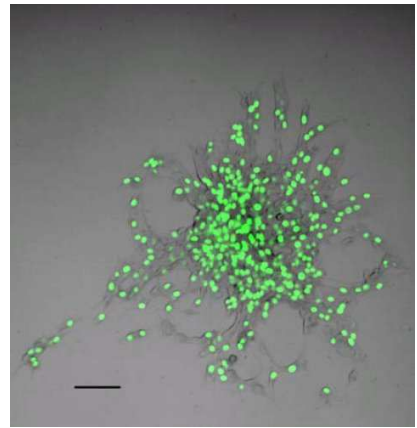} 
\end{center}
\caption{Depicted is an overlaid image of human U87 brain tumor cells
which were stably transfected with a Green Fluorescent Protein (EGFP)
Histone 2B marker for nuclei. The image is taken from a central
cross-section of the MTS cultured in a three-dimensional extracellular
matrix (Matrigel, Becton Dickinson, USA) environment in vitro. Note
the chain-like invasive patterns. The image was taken one day post
transferring the MTS from liquid medium to Matrigel (scale bar = 100
um).}
\label{exp_tumor}
\end{figure}

We can define a {\em small} parameter $\varepsilon^2=\mu_M/\mu_U$. As
the diffusion coefficient of $M$ is so small, the evolution of the
density field $M$ is slow in time, and therefore, random fluctuations
in its initial condition remain at late times. Hence, the underlying
medium is {\em quenched} from the perspective of the tumor
cells. Moreover, these fluctuations take place on {\em fast} length
scales, namely, we can write $\mu_U(M)\equiv \bar\mu_U({\bf
x}/\varepsilon)$~\cite{nota_eps}.  There are many studies devoted to
the propagation of fronts in heterogeneous media (as is the case here
for late times from the point of view of the tumor
cells)~\cite{xin}. Thus, it can be shown that, to leading order in
$\varepsilon$, Eq.~(\ref{prolif}) can be assumed homogeneous. Namely,
we can make the substitution
\begin{equation}
\bar\mu_U({\bf x}/\varepsilon)\rightarrow \left\langle 
\frac{1}{\bar\mu_U}\right\rangle^{-1}\equiv \mu_U=\textrm{constant}
+O(\varepsilon),
\label{homogen}
\end{equation}
where $\langle\ldots\rangle$ denotes the average over a region of
length $l$ much greater than the characteristic length scale of the
quenched fluctuations of $M$. This means that to lowest order we can
assume a constant diffusion coefficient for $U$ and that
Eq.~(\ref{prolif}) becomes the well-known Fisher
equation~\cite{fisher}. Fisher's equation admits planar traveling
front solutions, with minimal wave speed $v_0$ given
by~\cite{speed-fisher}
\begin{equation}
v_0=2(\mu_U \lambda_U K)^{1/2}.
\label{V_0}
\end{equation}
Moreover, any deviation from the planar front (or circular for
two-dimensional tumors) damps out, so cell proliferation by itself
cannot generate the branches observed in our experiments (see
Fig.~\ref{exp_tumor}).
Equation~(\ref{V_0}) provides the mean velocity of the tumor whenever
proliferation is the only mechanism of tumor growth. However,
chemotaxis drives tumor cells faster than proliferation itself, so
$v_0$ is a small quantity.  Thus, we can infer that cell proliferation
is a long time process and consequently $\lambda_M \approx 0$ and
$\lambda_U\approx 0$. We, therefore, assume that tumor cell
proliferation is much slower than the chemotactically induced tumor
cell growth (see Sec.~\ref{sec:hetero}).

Equation~(\ref{homogen}) is only valid to lowest order in
$\varepsilon$. Corrections to the leading behaviour of $\mu_U$ provide
also corrections to the velocity $v_0$. It can be shown
that~\cite{xin}
\begin{equation}
v_0=2(\mu_U \lambda_U K)^{1/2}(1+\xi^{1/2}),
\label{V_0_noise}
\end{equation}
where $\xi$ is obtained from the expansion of $\mu_U(M)$ to first
order in $\varepsilon$, and can be understood as a quenched noise
term, {\em i.e.,} a time independent random
function~\cite{xin}. Curvature corrections to Eq.~(\ref{V_0_noise})
give the so-called quenched Kardar-Parisi-Zhang
equation~\cite{barabasi}. Thus, the heterogeneity of the matrigel
medium will produce rough tumor interfaces. As we will see below, some
tumor fluctuations (large length scales) are amplified by chemotaxis,
so they act as {\em initial seeds} for invasive branches.

\subsubsection{$\mu_M\gg\mu_U$}

Despite the fact that the branching morphology in Fig.~\ref{exp_tumor}
has been obtained in the case $\mu_M\ll\mu_U$, for completeness, we
include in this section the opposite limit as well.

\begin{figure}[!hbt] 
\begin{center}
\includegraphics[width=0.4\textwidth,clip=]{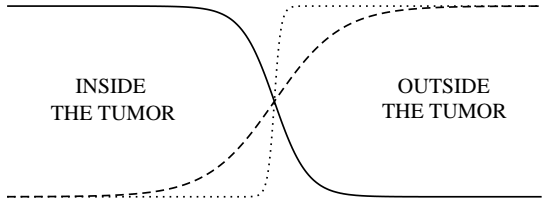} 
\end{center}
\caption{$M$ density field for the two limiting cases: $\mu_M\ll\mu_U$
(dotted line) and $\mu_M\gg\mu_U$ (dashed line). The solid line
represents the density field of tumor cells $U$.}
\label{limits}
\end{figure}

In this limit the diffusion coefficient of $M$ is so large that any
fluctuation of $M$ is rapidly damped out. In this case, we cannot
clearly separate the regions where $M$ takes its limiting values ($0$
and $M_0$), as was possible to do in the limit $\mu_M\ll\mu_U$ (see
Fig.~\ref{limits}). In this case we have to deal with the full
nonlinear equation that includes the dependence of $\mu_U$ on $M$.
Hence, the specific form of the diffusion coefficient $\mu_U(M)$ is
required in order to fully understand the evolution of the
system. M\"uller and van Saarloos~\cite{saarloos} have studied the
specific case in which (in our notation)
$\mu_U(K-\lambda_M/\lambda_UU)\sim U^k$, with $k>0$. In such case, the
gradient of the tumor field density $U$ at the boundary of the tumor
is discontinuous. This could be checked experimentally in order to
determine the qualitative form of the diffusion coefficient
$\mu_U(M)$.

\subsection{Growth due to heterotype chemoattraction}
\label{sec:hetero}

In this paper we are interested in the case where the nutrient medium
$M$ diffuses slowly, and so the homogenization given by
Eq.~(\ref{homogen}) can be assumed for all diffusion coefficients and
chemotactic sensitivities. We, therefore, drop any dependence of these
quantities on $M$. In what follows we restrict ourselves to a
two-dimensional study, with no loss of generality (the
three-dimensional analysis can be carried out as well).

As we have shown, cell proliferation cannot by itself provide invasive
behaviour. The next mechanism that we must include in order to
understand cell invasion is heterotype chemoattraction, where the
attractant molecules are provided externally to the tumor. They
diffuse rapidly until they reach the tumor boundary and then two
independent events take place: the heterotype chemoattractant is
degraded by the tumor cells and the tumor cells are ({\em
chemotactically}) drifted to higher heterotype chemoattractant
concentration gradients.

The consumption rate of the heterotype chemoattractant, $R_Q(Q,U)$,
cannot be arbitrarily large as it saturates for large values of
$Q$. This assumption is based on the concept that each tumor cell
carries a finite number of $Q$-uptaking cell receptors, which in turn
determine the cell's maximum uptake rate. Besides, it is also a
growing function of the tumor cell density, $U$. We do not need to
specify the precise mathematical form of $R_Q$ at this point, but
taking into account the above assumptions, we can write, without loss
of generality,
\begin{equation}
R_Q(Q,U)=U^\gamma f(Q),
\end{equation}
with $\gamma$ a positive constant.

In summary, the evolution equations in this case are
\begin{equation}
\partial_tU= \mu_{U} \nabla^2U- \nabla(U \chi_{Q}(Q) \nabla Q),
\label{u+q}
\end{equation}
and 
\begin{equation}
\partial_t Q= \mu_Q \nabla^2 Q -a_Q U^\gamma f(Q),
\label{qgamma}
\end{equation}
where we have assumed that tumor cell proliferation is negligible
compared to chemotaxis, so we can set $\lambda_M=\lambda_U=v_0=0$. In
this limit the dynamics of $M$ is uncoupled from that of $Q$ and $U$.

Despite the fact that Eqs.~(\ref{u+q}) and~(\ref{qgamma}) are highly
nonlinear, due to the functions $\chi_Q(Q)$ and $f(Q)$, we can obtain
useful information by properly rescaling space, time and both field
densities $Q$ and $U$. Thus, considering an initial tumor
concentration $U_0$ located in a bounded region of the system and a
replenished source of heterotype chemoattractant modeled by
Eqs.~(\ref{bc2d}), we define:
\begin{eqnarray}
{\bf x}^\prime&=&{\bf x}\left(\frac{a_Q}{\mu_Q}\right)^{1/2}, \\
t^\prime&=&ta_Q, \\
u&=&U/U_0,\\
q&=&Q/Q_0.
\end{eqnarray}
Eqs.~(\ref{u+q}) and~(\ref{qgamma}) can be written (we drop primes for
clarity) as follows
\begin{eqnarray}
\partial_tu &=& \mu_U/\mu_Q \nabla^2u- \nu Q_0/\mu_Q
\nabla(u \bar\chi(q) \nabla q),
\label{u+q2}
\\
\partial_t q &=& \nabla^2 q - u^\gamma U_0^\gamma f(Q_0q)/Q_0,
\label{qgamma2}
\end{eqnarray}
where $\bar\chi(q)$ is a dimensionless version of $\chi_Q (Q)$ and
$\nu$ is defined through the relation $\nu\equiv \chi_Q(Q)/
\bar\chi(q)$.

Note that $\mu_Q$ is the fastest diffusivity in the problem, so we can
define an small parameter $\epsilon=\mu_U/\mu_Q$. Moreover, we also
assume that the {\em cross} diffusion coefficient $\nu Q_0$ is smaller
than $\mu_Q$~\cite{cross-diffusion}. With these considerations
Eq.~(\ref{u+q2}) becomes
\begin{equation}
\partial_tu=\epsilon \nabla^2u- \rho\epsilon\nabla(u \bar\chi(q) \nabla q),
\label{u+q3}
\end{equation}
with $\rho=\nu Q_0/\mu_U$. Typically, the tumor field will be constant
almost everywhere except in a narrow region (boundary
layer~\cite{nayfeh}). This region defines an {\em interface} between
the inside and the outside of the tumor.

In fact, if we take the limit $\epsilon\rightarrow 0$ (the so-called
{\em outer} limit~\cite{nayfeh}) in Eq.~(\ref{u+q3}), we have
$\partial_tu=0$, and
\begin{equation}
u=\Big\{
\begin{array}{lc}
1 & \textrm{inside the tumor,}\\
0 & \textrm{outside the tumor.}\\
\end{array}
\end{equation}
One can see that in the outer limit the mathematical analysis of the
problem is much simpler as Eq.~(\ref{qgamma2}) reduces to the
equations
\begin{equation}
\partial_t q= \Big\{
\begin{array}{ll}
\nabla^2 q - U_0^\gamma f(Q_0q)/Q_0&\textrm{ inside the tumor,}\\
 \nabla^2 q& \textrm{ outside the tumor.}
\end{array}
\label{outer}
\end{equation}

In the case $\epsilon \neq 0$ (the so-called {\em inner}
limit~\cite{nayfeh}) we need to proceed with care: in order to solve
Eq.~(\ref{u+q2}) for $u$ and Eq.~(\ref{qgamma2}) for $q$, we need the
behavior of $q$ exactly at the tumor boundary layer (interface).  With
this in mind, we define a new local set of curvilinear coordinates: a
coordinate $n$ normal to the tumor interface and a coordinate $s$
tangential to it. Elementary computations (see Ref.~\cite{fife}) give
us the formulas for converting derivatives with respect to ${\bf x}$
to derivatives with respect to the new coordinates $(n,s)$:
\begin{equation}
\nabla^2=\partial_{nn}+\tilde \kappa\partial_n+\Delta_s,
\label{eq:nabla}
\end{equation}
where $\tilde \kappa$ is the local dimensionless curvature of the
tumor interface and $\Delta_s$ is the surface Laplacian~\cite{fife}.
Similarly, we can write
\begin{equation}
\partial_t=\partial_t-\tilde v\partial_n+s_t\partial_s,
\label{eq:time}
\end{equation}
where $\tilde v$ is the normal (dimensionless) velocity of the
interface. We also need the following derivative
\begin{equation}
\nabla(u \bar\chi(q) \nabla q)
=
\partial_n(u \bar\chi(q) \partial_n q)
+
\tilde \kappa (u \bar\chi(q) \partial_n q)
+
\nabla_s(u \bar\chi(q) \nabla_s q).
\label{eq:chemo-change}
\end{equation}
Our aim now is to find solutions for $q$ and $u$ in the tumor
interface. We start by rescaling the normal coordinate $n$, in terms
of the fast variable $\eta=n/\epsilon$, and time in terms of
$\tau=t/\epsilon$. We denote by $\tilde{u}$ and $\tilde{q}$ the {\em
inner} fields, and expand them, $\bar \chi(\tilde q)$ and $\tilde v$
in a power series of $\epsilon$~\cite{almgren}:
\begin{eqnarray}
\tilde{u}(\eta,s,t)&=&\tilde{u}_0+\epsilon \tilde{u}_1+ \ldots,\\
\tilde{q}(\eta,s,t)&=&\tilde{q}_0+\epsilon \tilde{q}_1+ \ldots,\\
\tilde v(\eta,s,t)&=&\tilde v_0+\epsilon \tilde v_1+ \ldots,
\\
\bar \chi(\tilde q)&=&\bar \chi(\tilde q_0)
+\bar \chi'(\tilde q_0)\epsilon \tilde q_1+ \ldots,
\label{chi-exp}
\end{eqnarray}
where the prime in Eq.~(\ref{chi-exp}) denotes a derivative with
respect to $\tilde q$.  Using Eqs.~(\ref{u+q3}), ~(\ref{eq:nabla}),
~(\ref{eq:time}) and~(\ref{eq:chemo-change}), the original system of
equations can be written as
\begin{eqnarray}
\partial_\tau \tilde{u}_0+ \epsilon \partial_\tau \tilde{u}_1
&=&        
\epsilon \tilde v_1\partial_\eta \tilde{u}_0 
+ \partial_{\eta\eta}\tilde{u}_0
\nonumber \\        
&+&
\epsilon\partial_{\eta\eta}\tilde{u}_1    
-\epsilon s_t \partial_s \tilde{u}_0
+\epsilon \tilde \kappa\partial_\eta \tilde{u}_0                   
\nonumber \\        
&-&
\rho\partial_\eta(\tilde{u}_0\bar\chi(\tilde{q}_0)\partial_\eta       
\tilde{q}_0)
-                           
\epsilon\rho\partial_\eta(\tilde{u}_1\bar\chi(\tilde{q}_0)\partial_\eta    
\tilde{q}_0)                                 
\nonumber \\
&-&
\epsilon\rho\partial_\eta(\tilde{u}_0\bar\chi(\tilde{q}_0)\partial_\eta   
\tilde{q}_1)\nonumber  -                          
\epsilon\rho\partial_\eta(\tilde{u}_0\bar\chi'(\tilde{q}_0)
\tilde{q}_1\partial_\eta \tilde{q}_0) 
\nonumber 
\\
&-& \epsilon\rho \tilde \kappa          
\tilde{u}_0\bar\chi(\tilde{q}_0)\partial_\eta \tilde{q}_0+O(\epsilon^2),   
\label{sing_u}
\\
\epsilon\partial_\tau \tilde{q}_0&=& 
\partial_{\eta\eta}\tilde{q}_0+\epsilon\partial_{\eta\eta}\tilde{q}_1     
+\epsilon \tilde \kappa\partial_\eta \tilde{q}_0 
+ O(\epsilon^2).           
\label{sing_q}
\end{eqnarray}

\subsubsection{Order $\epsilon^0$}

We start by solving Eqs.~(\ref{sing_u}) and~(\ref{sing_q}) to order
$\epsilon^0$.  As we stated above, tumor cell proliferation is
negligible when compared to chemotaxis, which means that $\tilde
v_0=0$. Otherwise we have to include the reaction term $R_M$ in the
dynamical equation for $U$ and then, to order $\epsilon^0$,
Eq.~(\ref{sing_u}) becomes:
\begin{eqnarray}
\partial_\tau \tilde{u}_0+ \epsilon \partial_\tau \tilde{u}_1
&=&        
\tilde v_0\partial_\eta \tilde{u}_0
+ \epsilon \tilde v_1\partial_\eta \tilde{u}_0 
+ \epsilon \tilde v_0\partial_\eta \tilde{u}_1 
+ \partial_{\eta\eta}\tilde{u}_0
\nonumber \\        
&+&
\epsilon\partial_{\eta\eta}\tilde{u}_1    
-\epsilon s_t \partial_s \tilde{u}_0
+\epsilon \tilde \kappa\partial_\eta \tilde{u}_0                   
\nonumber \\        
&-&
\rho\partial_\eta(\tilde{u}_0\bar\chi(\tilde{q}_0)\partial_\eta       
\tilde{q}_0)
-                           
\epsilon\rho\partial_\eta(\tilde{u}_1\bar\chi(\tilde{q}_0)\partial_\eta    
\tilde{q}_0)                                 
\nonumber \\
&-&
\epsilon\rho\partial_\eta(\tilde{u}_0\bar\chi(\tilde{q}_0)\partial_\eta   
\tilde{q}_1)\nonumber  -                          
\epsilon\rho\partial_\eta(\tilde{u}_0\bar\chi^\prime(\tilde{q}_0)
\tilde{q}_1\partial_\eta \tilde{q}_0) 
\nonumber 
\\
&-& \epsilon\rho \tilde \kappa          
\tilde{u}_0\bar\chi(\tilde{q}_0)\partial_\eta \tilde{q}_0
\nonumber \\        
&+&
\epsilon \frac{\lambda_U K}{a_Q} 
\left(1 - \frac{\lambda_M U_0}{\lambda_U K} \tilde{u}_0 \right)\tilde{u}_0
+O(\epsilon^2).   
\label{sing_u_new}
\end{eqnarray}
Notice that in the absence of chemotaxis Eq.~(\ref{sing_u_new})
becomes Fisher's equation in the set of coordinates ($\eta,s$) and
with time given by $\tau$.

{From} Eq.~(\ref{sing_q}) we find
\begin{equation}
\partial_{\eta\eta}\tilde{q}_0=0.
\end{equation}
We know that $\tilde{q}_0$ must be bounded at $\eta=\pm \infty$,
corresponding to the inside and outside of the tumor in the scaled
variable $\eta$. This implies $\partial_\eta \tilde{q}_0=0$ and,
therefore, $\tilde{q}_0=\mathrm{constant}$. Substituting this into
Eq.~(\ref{sing_u}), we obtain
\begin{equation}
\partial_\tau \tilde{u}_0=\partial_{\eta\eta}\tilde{u}_0,
\end{equation}
which states the fact that, to first order, the tumor cells simply
diffuse. Note that to this order $\tilde {u}_0$ does not depend on
$s$, so that $\partial_s \tilde u_0=0$, due to the boundary conditions
on $u$.

\subsubsection{Order $\epsilon^1$}

To next order in $\epsilon$ and {from} Eq.~(\ref{sing_q}) we find,
\begin{equation}
\partial_{\eta\eta}\tilde{q}_1=0 \Rightarrow \partial_\eta \tilde{q}_1
=\textrm{constant},
\end{equation}
due to the boundary conditions on $q$.

The next order equation for $\tilde{u}_1$ is given by
\begin{equation}
\partial_\tau \tilde{u}_1
= \tilde v_1\partial_\eta \tilde{u}_0+           
\partial_{\eta\eta}\tilde{u}_1 +\tilde \kappa\partial_\eta \tilde{u}_0        
-\rho(\partial_\eta \tilde{u}_0)\bar\chi(\tilde{q}_0)\partial_\eta      
\tilde{q}_1.                                 
\end{equation}
The Fredholm alternative (or solvability condition) for $\tilde{u}_1$
provides~\cite{almgren}
\begin{equation}
\tilde v_1=-\tilde \kappa+\rho B\partial_\eta \tilde{q}_1,
\end{equation}
where
\begin{equation}
B=\frac{\int_{-\infty}^\infty d\eta (\partial_\eta
\tilde{u}_0)^2\bar\chi(\tilde{q}_0)}{\int_{-\infty}^\infty d\eta
(\partial_\eta \tilde{u}_0)^2}.
\end{equation}
Matching the inner and outer expansions yields the normal velocity of
the tumor in the original dimensions~\cite{almgren}:
\begin{equation}
v_n=-\mu_U\kappa+\nu B\nabla Q\cdot {\bf n},
\label{velocity}
\end{equation}
where ${\bf n}$ is a unit vector normal to the tumor interface and
directed away from the tumor.

Chemotactically induced tumor growth requires $\nabla Q\cdot {\bf n}$
to be a positive quantity, so that $v_n$ is positive. This can be
achieved whenever the heterotype chemoattractant concentration grows
as we move away from the tumor surface. Before performing the analysis
of Eqs.~(\ref{outer}) and~(\ref{velocity}) to check this requirement,
we consider two interesting features related to
Eq.~(\ref{velocity}). First of all, in the absence of chemotaxis and
proliferation, for an initially circular tumor of radius $R_0$,
Eq.~(\ref{velocity}) reduces to
\begin{equation}
\frac{dR}{dt}=-\frac{\mu_U}{R},
\end{equation}
which gives the diffusive behavior $R(t)=\sqrt{R_0^2 -2 \mu_Ut}$.
This means that the mean velocity of the tumor due to diffusion decays
as $t^{-1/2}$.  At this stage, and for an initially circular tumor,
one could now perform the following analysis: (i) suppose the tumour
continues to grow as a two-dimensional disc and (ii) perturb the
boundary and study the development of instabilities. This is out of
the scope of this paper and will be addressed in future work.
Secondly, for a general tumor front geometry, Eq.~(\ref{velocity})
provides a critical curvature, (reminiscent of the classical Greenspan
model~\cite{greenspan})
\begin{equation}
\kappa_c=\frac{\nu B}{\mu_U}\nabla Q\cdot{\bf n},
\label{c-crit}
\end{equation}
such that for $\kappa<\kappa_c$ the tumor interface has locally a
positive velocity, for $\kappa>\kappa_c$ the tumor interface has
locally a negative velocity and for $\kappa=\kappa_c$ the tumor
interface has locally vanishing velocity. Note that this is precisely
the reason of the dynamical instability that guarantees the
development of invasive branches. Tumor invasion will take place on
the tumor surface wherever the local curvature is below the critical
value $\kappa_c$, given by Eq.~(\ref{c-crit}).  Our result agrees with
the specific case considered in Ref.~\cite{libro-sleeman}.

As a final remark concerning Eq.~(\ref{velocity}), note that it
resembles the equation for the velocity of a solidification
front~\cite{langer,karma}. In that case, the local normal velocity of
the front depends on the local curvature, $\kappa$ as above, but on
the value of the field (temperature) at the boundary, instead of the
value of the gradient, $\nabla Q$, at the tumor interface. This is to
be expected as in the tumor picture it is the heterotype
chemoattractant that is driving the dynamics, and in the case of a
solidification front, the dynamics of the front is linked to the
temperature field~\cite{langer,karma}.  The branching behaviour of the
tumor also resembles the fingering instability of the Hele-Shaw
problem~\cite{hele-shaw}

\begin{figure}[!th] 
\begin{center}
\includegraphics[width=3cm,clip=]{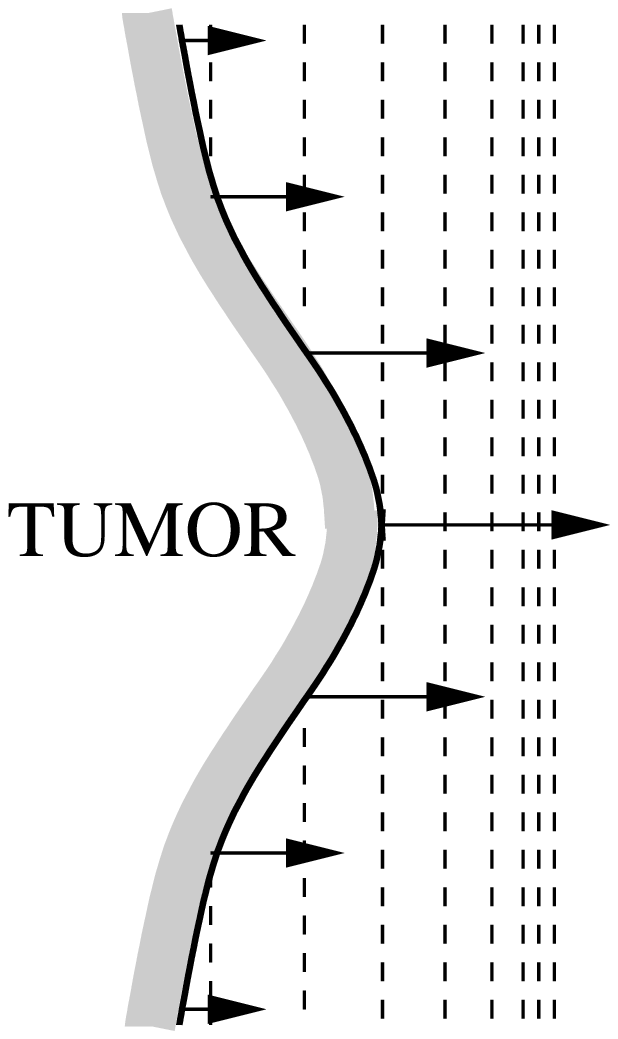} 
\end{center}
\caption{Sketch of the heterotype chemoattractant effect on tumor cell
invasive drift. Dashed lines represent level sets of $Q$ and arrows
represent the local normal velocity of the tumor cells due to
chemotaxis (the length of the arrows is proportional to the normal velocity).}
\label{instability}
\end{figure}

We now turn to the requirement on the value of $\nabla Q\cdot {\bf
n}$. In order to check that, indeed, $\nabla Q\cdot{\bf n}$ (or
equivalently $\nabla q\cdot {\bf n}$) is a positive quantity at the
tumor boundary, we need to solve Eq.~(\ref{outer}) supplemented by
Eq.~(\ref{velocity}), with initial condition,
\begin{equation}
q(x,y,t=0)=\delta(x-L_x),
\label{q-init}
\end{equation}
and boundary conditions 
\begin{eqnarray}
q(x=0,y,t)=0, \\
q(x=L_x,y,t)=1. 
\end{eqnarray}
These equations cannot be solved in general without the precise form
of the function $f(Q)$. Nevertheless, we may assume that, as tumor
cells degrade the heterotype chemoattractant, the concentration of $q$
inside the tumor is small enough, and we can approximate $f(Q)$ by a
linear function $f(Q_0q)\simeq \delta q$. We now proceed by solving
Eq.~(\ref{outer}) inside and outside the tumor (we denote the
solutions, respectively, by $q^-$ and $q^+$) and matching both
solutions at the boundary, for an initially {\em flat} tumor, {\em
i.e.,} a tumor for which the radius of curvature is much smaller than
the system lateral size $L_x$~\cite{nota_flat}. We first notice that
$q^+$ is invariant under the transformation group $(L_x-x)\rightarrow
\epsilon(L_x-x)$, $t\rightarrow \epsilon^2t$ and $q^+\rightarrow
\epsilon^0q^+$.  Similarly, $e^{\delta t}q^-$ is invariant under the
group, $x\rightarrow \epsilon x$, $t\rightarrow \epsilon^2t$ and
$q^-\rightarrow \epsilon^0q^-$.  Thus, it can be straightforwardly
seen that~\cite{grindrod}
\begin{eqnarray} 
q^-(x,y,t)&=&Ce^{-\delta
t}\int_0^{x/\sqrt{t}}\, d\xi \; e^{-\xi^2/4},\\
q^+(x,y,t)&=&1-A\int_0^{(L_x-x)/\sqrt{t}}\, d\xi \; e^{-\xi^2/4},
\end{eqnarray}
where $A$ and $C$ are positive constants that can be determined by
continuity of the solutions at the tumor boundary ${\bf x}={\bf
x_0}(s,t)$. As we are interested in the gradient of the $Q$ density
field, we find
\begin{eqnarray}
\partial_xq^- (x,y,t)&=&\frac{C}{\sqrt{t}}e^{-\delta t}e^{-x^2/4t},\\
\partial_xq^+ (x,y,t)&=&\frac{A}{\sqrt{t}}e^{-(L_x-x)^2/4t}.
\label{gradq+}
\end{eqnarray}
Note that Eq.~(\ref{gradq+}) states that the gradient of $Q$ is a
positive function, and so the tumor velocity is increased by
chemotaxis as we had anticipated. Moreover, the larger the distance
from $x$ to the tumor is, the larger the value of $\partial_xq^+$
becomes and the larger the value of the velocity front is (see
Fig.~\ref{instability}). This is indeed, the reason why small
fluctuations on the tumor surface become emerging invasive branches.
We will see in Sec.~\ref{sec:num} that this chemotactically enhanced
velocity is also obtained numerically.

\subsection{Growth due to homotype chemoattraction} 
\label{sec:homo}

Finally, we consider the role of homotype chemoattractants.
Representing protein growth factors, these chemoattractants are
produced and internalized~\cite{internal}, or (for the purposes here),
consumed by the tumor cells that move towards their positive
gradients, in a similar fashion as in the heterotype case. Yet, there
is no wide time scale separation between tumor and homotype
chemoattractant dynamics. Thus, we cannot, in general, neglect tumor
cell proliferation when analyzing homotype chemoattractant
dynamics. This means that the equations in this case are
Eqs.~(\ref{basic-m}),~(\ref{basic-c}) and~(\ref{basic-u}) with
$\chi_Q=0$. Performing a similar analysis to that of
Sec.~\ref{sec:hetero} above we find:
\begin{equation}
v_n=2(\mu_U\lambda_U K)^{1/2}-\mu_U\kappa+\nu^\prime 
B^\prime\nabla C\cdot {\bf n}.
\label{velocity_C}
\end{equation}
Despite the fact that Eq.~(\ref{velocity_C}) is equivalent to
Eq.~(\ref{velocity}), the main differences between the evolution of
$Q$ and $C$ are due to the behaviour of both density fields away from
the tumor interface, namely, due to the production term $R_C^{(p)}$
(that is absent in the dynamical equation for $Q$) and the different
boundary conditions (no external source for $C$).

Following the same steps as those carried out in
Sec.~{\ref{sec:hetero}}, we find that the {\em outer} limit yields the
following equation
\begin{equation}
\partial_t c= \Big\{
\begin{array}{ll}
\nabla^2 c +r^{(p)}(0,1)-r^{(d)}(c,1)&\textrm{ inside the tumor,}\\
 \nabla^2 c& \textrm{ outside the tumor,}
\end{array}
\label{outer_C}
\end{equation}
with $r^{(p)}$ and $r^{(d)}$ scaled (dimensionless) versions of
$R_C^{(p)}$ and $R_C^{(d)}$, respectively. Note that we have
considered $m=0$ where $u=1$. The conservation relation given by
Eq.~(\ref{M_U}) still holds in this case so, without loss of
generality, we reduce the system of
equations~(\ref{basic-m}),~(\ref{basic-c}) and~(\ref{basic-u}) to
Eq.~(\ref{basic-c}) and Eq.~(\ref{basic-u}), with $\chi_Q=0$ and $M$
given by Eq.~(\ref{M_U}).

Just as we did in Sec.~{\ref{sec:hetero}, we must now compute the sign
of $\nabla C\cdot{\bf n}$ in order to determine whether or not
homotype chemotaxis increases the velocity of the tumor boundary, and
if it can generate a dynamical instability leading to a tumor
branching morphology. We cannot in general find a transformation group
under which Eq.~(\ref{outer_C}) is invariant.  This means that we
cannot study homotype chemotaxis with the tools of the previous
section.  However, we can analyze homotype chemoattraction by means of
its homogeneous and steady state solutions, based on generic
assumptions regarding the reaction terms $R_C^{(p)}(M,U)$ and
$R_C^{(d)}(C,U)$.

The homogeneous, steady state solutions
(nullclines~\cite{strogatz}) of equations~(\ref{prolif})
and~(\ref{basic-c}) are given by the solutions of
\begin{eqnarray}
R_M(K-\lambda_M/\lambda_U U, U)&=&0, 
\label{nullcline1} \\ 
R_C^{(p)}(K-\lambda_M/\lambda_U U, U) &=& R_C^{(d)}(C,U).
\label{nullcline2}
\end{eqnarray}
These nullclines yield the fixed points $U_1=0$ and
$U_2=\lambda_UK/\lambda_M$, and $C_1$ and $C_2$ given implicitly by
Eq.~(\ref{nullcline2}). We must distinguish between the inside and the
outside of the tumor when analyzing tumor growth due to homotype
chemotaxis. Outside the tumor $U_1=0$ and there is no production of
$C$ (as there are no tumor cells). This implies that the corresponding
fixed point value of $C$ is then $C_1=0$. On the other hand, inside
the tumor, Eq.~(\ref{nullcline2}) reflects the fact that there is a
balance between production and consumption of $C$. This means that the
value of the density field $C$ reaches the fixed point $C_2$, which is
a constant (equilibrium) value. This simple picture holds even far
enough from the tumor boundary (diffusion tends to spread these
uniform concentration phases). Therefore, let us consider a point in
the $C-U$ plane where $U=U_P$ and $C=0$ (point P in
Fig.~\ref{pulse}a), namely, a point outside the tumor boundary, with
$U$ so small that there has been no previous secretion of homotype
chemoattractant $C$. The concentration $C$ eventually grows (as $U$
increases due to proliferation wherever $M \neq 0$) because the slope
given by
\begin{equation}
\frac{dC}{dU}=
\frac{R_C^{(p)}(K-\lambda_M/\lambda_U U,U)
-
R_C^{(d)}(C,U)}{R_M(K-\lambda_M/\lambda_U U,U)}, 
\label{dcdu}
\end{equation} 
is positive~\cite{note_null_1}.  Then, the slope
decreases~\cite{note_null_2} until it reaches the nullcline where the
slope is $0$ (point Q in Fig.~\ref{pulse}a). Finally, the curve
approaches the stable point $(U_2,C_2)$ with infinite slope (point R
in Fig.~\ref{pulse}a). Thus, $C$ grows from the outside of the tumor,
reaches a maximum value and then decreases to a constant value $C_2$
inside the tumor. This can be schematically seen in
Fig.~\ref{pulse}b). In other words, the density field $C$ tends to
grow as $U$ decreases namely, as we move outside of the tumor from
inside. But, as we have shown, outside the tumor and far enough from
it $C$ tends to $0$. This means that $C$ also tends to grow as we move
inside of the tumor from the outside.  Consequently, $C$ behaves as a
pulse from the constant value $C_2$ inside the tumor to $C_1=0$
outside the tumor. This qualitative picture is sketched in
Fig.~\ref{pulse}, where we have shown that the nullcline given by
Eq.~(\ref{nullcline2}) can have three different qualitative behaviours
$N_1$, $N_2$ and $N_3$, that are plotted as well.  This pulse-like
structure for the density profile of $C$ agrees well with the
intuitive picture provided in Sec.~\ref{sec:C}. It also explains why
tumor cells are chemotactically guided by the gradient of $C$. These
facts will be numerically confirmed in Sec.~\ref{sec:num} below. We
conclude as follows: the density profile of $C$ grows at the tumor
boundary, which implies that the local normal velocity of the tumor
surface increases due to the presence of a positive homotype
chemoattractant gradient.
\begin{figure}[!hbt] 
\begin{center}
\includegraphics[width=0.4\textwidth,clip=]{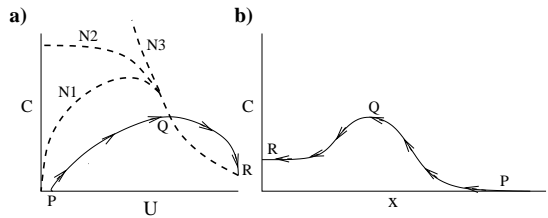} 
\end{center}
\caption{a) Qualitative behaviour of the {\em trajectory} of the
density field $C$ in the {\em phase-plane}. The arrows indicate the
direction of time evolution. Curves $N1$-$N3$ are the three types of
nullclines that can be expected for our system. b) Same trajectory of
the density field $C$ with respect to the spatial coordinate
$x$. Again, arrows indicate time evolution.}
\label{pulse}
\end{figure}

\section{Numerical study}
\label{sec:num}

In Sec.~\ref{sec:analy} we have presented a general analytical
framework to study tumor growth (proliferation and chemotactic
invasion).  The previous analysis has been carried out by means of
several assumptions and limits, ({\em e.g.,} $\nu Q_0 \ll \mu_Q$,
$\lambda_M \approx 0, \lambda_U \approx 0$), that have not been fully
justified.  In this section we provide numerical simulations that
check the validity of those assumptions and limits.  We, thus,
numerically solve the main differential equations of
Sec.~\ref{sec:model} following a similar organization to that of
Sec.~\ref{sec:analy}. It is clear that in order to carry out a
numerical study we must specify the detailed mathematical form of the
reaction terms in our equations~(\ref{basic-m})-~(\ref{basic-u}).
These reaction terms are chosen following the spirit of the oncology
concept previously reported in Refs.~\cite{prolif,physicaa}.

\subsection{Growth due to cell proliferation}
\label{sec:num-M}

In this section, we check the validity of Eq.~(\ref{M_U}) as an
approximation to Eq.~(\ref{M_U_1}). Thus, we have integrated
Eqs.~(\ref{basic-m}) and~(\ref{basic-u}) with $\chi_Q=\chi_C=0$ and
$R_M(U,M)=MU$ and, independently, Eq.~(\ref{basic-u}) with
$M=K-\lambda_M/\lambda_UU$.

\subsubsection{One-dimensional results}

Initially, we place a {\em one dimensional tumor} such that
$U(x,t=0)=1$ for $0<x<L/2$ and $U(x,t=0)=0$ elsewhere. This implies
that at the initial time $M(x,t=0)=1-U(x,t=0)$.  We have numerically
solved Eqs.~(\ref{basic-m}) and~(\ref{basic-u}) with $\chi_Q=\chi_C=0$
and $R_M(U,M)=MU$ and, independently, Eq.~(\ref{basic-u}) with
$M=K-\lambda_M/\lambda_UU$.  The lattice separation has been taken
$dx=0.5$, the lattice size $L_x=128$, the diffusion coefficients
$\mu_U=0.01$ and $\mu_M=0.0$, the reaction rates
$\lambda_M=\lambda_U=0.1$, the time step $\epsilon_t=0.005$, the
initial time $t_0=\epsilon_t$ and the final time $t_f=
50000\epsilon_t$.

As can be seen in Fig.~\ref{fisher-fig}, after a small transient time,
the approximation is accurate enough. As we mentioned in
Sec.~\ref{sec:analy}, the shape of the tumor front changes slightly.
\begin{figure}[!hbt] 
\begin{center}
\includegraphics[width=0.4\textwidth,clip=]{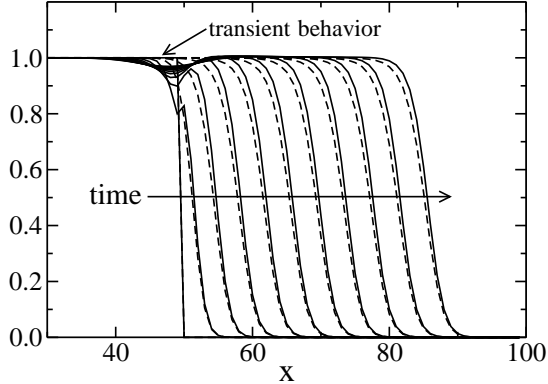} 
\end{center}
\caption{One dimensional tumor proliferation. Solid lines represent
the $U$ density field for the model given by Eqs.~(\ref{basic-m})
and~(\ref{basic-u}). Dashed lines represent $U$ for the solution of
Fisher's equation, {\em i.e.,} Eq.~(\ref{basic-u}) with
$M=K-\lambda_M/\lambda_UU$.}
\label{fisher-fig}
\end{figure}

\subsubsection{Two-dimensional results}

We now consider tumor growth due to proliferation in a two-dimensional
setting with $\chi_Q=\chi_C=0$. Initially, the matrigel density $M$ is
a random distribution to reflect the fact that it is an heterogeneous
medium. As we are interested in a slowly varying nutrient, we choose
$\varepsilon\equiv \mu_M/\bar\mu_U=0.01$, where $\bar\mu_U$ is the
maximum attainable value of $\mu_U(M)$. We assume that this value of
$\mu_U$ corresponds to the value $M=0$. The confinement due to the
matrigel is unlimited, namely, for large concentrations of $M$, tumor
cells can no longer diffuse. Hence, we take the following diffusion
coefficient
\begin{equation}
\mu_U(M)=\frac{\bar\mu_U}{1+M/M_{\rm th}},
\end{equation}
where $M_{\rm th}$ is a reference threshold concentration.                    

We have solved Eqs.~(\ref{basic-m}) and~(\ref{basic-u}) with the
following initial conditions: at time $t=0$ we place a circular tumor
centered at $(L_x/2,L_y/2)$ of radius $L_x/4$ and surrounded by a
heterogeneous nutrient substrate $M$. Thus, $M(x,y,t=0)=0$ inside the
initial circular tumor and $M(x,y,t=0)=1+\xi(x,y)$ elsewhere, with
$\xi(x,y)$ a random Gaussian distribution with zero mean and variance
$0.2$, that encodes the initial heterogeneities of the matrigel.  The
lattice separation has been taken $dx=dy=0.5$, the lattice size
$L_x=L_y=128$, the diffusion coefficients $\bar \mu_U=0.01$ and
$\mu_M=0.0$, the reaction rates $\lambda_M=\lambda_U=1.5$, the time
step $\epsilon_t=0.005$, the initial time $t_0=\epsilon_t$, the final
time $t_f= 5000\epsilon_t$, and the intermediate times
$(1000,2000,3000,4000)\epsilon_t$.

Figure~\ref{quenched} displays the time evolution of the tumor
surface.  Notice how the tumor conserves during its evolution its
initial circular shape but develops a rough interface with the
matrigel.
\begin{figure}[!hbt] 
\begin{center}
\includegraphics[width=0.2\textwidth,clip=]{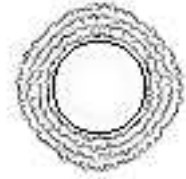} 
\end{center}
\caption{Numerical simulation of an initially circular tumor embedded
in a matrigel medium $M$ with slow dynamics ($\mu_M \ll \bar
\mu_U$). Different curves represent different times with the initial
time $t_0$ corresponding to the inner perfect circle.}
\label{quenched}
\end{figure}

\subsection{Growth due to heterotype chemoattraction}
\label{sec:num-Q}

The experimental branches of Fig.~\ref{exp_tumor} have two
characteristic lengths, namely, their width and their height with
respect to the main tumor substrate. Section~\ref{sec:hetero} was
devoted to determine the conditions that trigger the formation of the
invasive branches. We know that the height of the branches depends in
a crucial way on the mathematical form of the reaction term $R_Q(Q,U)$
of Eq.~(\ref{basic-q}), namely, the height depends on the
concentration thresholds associated with the bio-chemical reaction of
$Q$ consumption by tumor cells~\cite{prolif}. Thus, we choose
\begin{equation}
a_Q R_Q(Q,U)= a_Q U\frac{Q}{b_Q+Q},
\label{rq}
\end{equation}
where $a_Q$ is the inverse of the time scale of consumption and $b_Q$
a characteristic heterotype chemoattractant concentration (threshold
value). Note that, in the notation of Eq.~(\ref{qgamma}) we have set
$\gamma=1$.

\subsubsection{Two-dimensional results}

We have integrated Eqs.~(\ref{basic-m}),~(\ref{basic-u})
and~(\ref{basic-q}) with $R_Q$ given by Eq.~(\ref{rq}) and
$\chi_C=0$. At the initial time we place a circular tumor at
$(L_x/2,L_y/2)$ with radius $L_x/4+\xi$, where $\xi$ is a Gaussian
random number with zero mean and variance $L_x/20$. This initial
condition mimics the effect of the slowly varying underlying substrate
(matrigel) as shown in the previous section~\ref{sec:num-M}.  The
lattice separation has been taken $dx=dy=0.5$, the lattice size
$L_x=L_y=128$, the diffusion coefficients $\mu_U=0.001$, $\mu_Q=10.0$
and $\mu_M=0.00001$, the reaction rates $\lambda_M=\lambda_U=0.0050$,
$a_Q=0.75$, $b_Q=0.5$, the chemotactic sensitivity $\chi_Q=2.0$, the
time step $\epsilon_t=0.005$, the initial time $t_0=\epsilon_t$ and
the final time $t_f=20000\epsilon_t$.

Figure~\ref{tumor_sim}a) displays the
numerically obtained tumor when the replenished source of $Q$ is
placed at $x=L_x$ ({\em i.e.,} the right hand side of the
lattice). Moreover, in Fig.~\ref{tumor_sim}b) we show the
cross-section of an invasive, chemotactically induced branch obtained
in the same simulation. Clearly, we can distinguish between the main
tumor spheroid and a given branch. Notice the good agreement with the
experimental results and with the qualitative analysis provided in
Sec.~\ref{sec:analy}.
\begin{figure}[!hbt] 
\begin{center}
\includegraphics[width=0.5\textwidth,clip=]{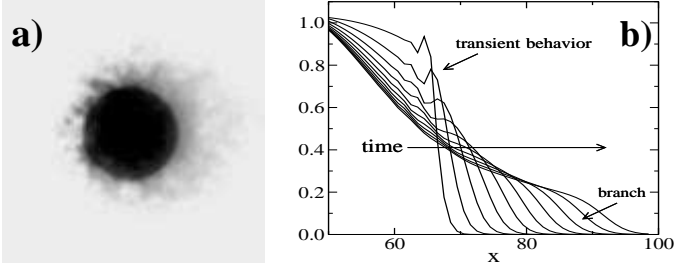} 
\end{center}
\caption{a) Numerical simulation of tumor branching induced by a
heterotype chemotactic source located at $x=L_x$. The plot represents
the tumor density field $U(x,y,t=20000\epsilon_t)$. b) Cross-section
of the tumor in panel a) for different times between $2000\epsilon_t$
and $20000\epsilon_t$.}
\label{tumor_sim}
\end{figure}

\subsection{Growth due to homotype chemoattraction} 
\label{sec:num-C}

The segregation and eventual degradation of the homotype
chemoattractant is limited, {\em i.e.,} the rates associated with both
processes cannot be arbitrarily large. Thus, following
Ref.~\cite{physicaa} we choose
\begin{equation}
\alpha_C 
R_C^{(p)}(U)= \alpha_C \frac{U}{\beta_{C}+U},
\label{rcp}
\end{equation}
\begin{equation}
a_C R_C^{(d)}(C,U)=a_C U\frac{C}{b_{C}+C}.
\label{rcd}
\end{equation}
The constants $\alpha_C$ and $a_C$ are the inverse of the
characteristic time scales of production and degradation,
respectively, and $\beta_C$ and $b_C$ are characteristic saturation
concentrations (for production and degradation, respectively). Note
the Eqs.~(\ref{rcd}) and~(\ref{rq}) have the same mathematical form.

\subsubsection{Two-dimensional results}

We have numerically integrated Eqs.~(\ref{basic-m}),~(\ref{basic-u})
and~(\ref{basic-c}) with $\chi_Q=0$, and $R_C^{(p)}$ and $R_C^{(d)}$
given by Eqs.~(\ref{rcp}) and~(\ref{rcd}), respectively.  The initial
conditions for $U(x,y,t=0)$ are the same as those chosen in the
previous section~\ref{sec:num-Q}. For the matrigel and the homotype
chemoattracctant, we have chosen $M(x,y,t=0)=1-U(x,y,t=0)$ and
$C(x,y,t=0)=0$, respectively.  The lattice separation has been taken
$dx=dy=0.5$, the lattice size $L_x=L_y=128$, the diffusion
coefficients $\mu_U=0.01$, $\mu_C=1.0$ and $\mu_M=0.0$,
the reaction rates
$\lambda_M=\lambda_U=0.1$, 
$a_C=1.75$, $b_C=0.1$,  $\alpha_C=\beta_C=1.0$, 
the chemotactic sensitivity $\chi_C=1.0$, 
the time step $\epsilon_t=0.005$,
the initial
time $t_0=\epsilon_t$ and the final time $t_f=50000\epsilon_t$.

Fig.~\ref{tumor_homo} displays the time evolution of the tumor profile
in the $x$ direction for times $10000\epsilon_t, 15000\epsilon_t,
20000\epsilon_t, 25000\epsilon_t$ and $30000\epsilon_t$.  The dotted line
in Fig.~\ref{tumor_homo} displays the $C$ profile at time
$t=10000\epsilon_t$, magnified by a factor of $4$.  In agreement with the
analysis presented in section~\ref{sec:homo}, the numerical results
show that (i) there is no emergence of chemotactically induced
branches, as was the case for the heterotype chemoattractant and (ii)
the mean speed of the tumor boundary increases due to $C$.  That is,
the tumor profile follows qualitatively the behaviour anticipated in
Sec.~\ref{sec:homo}.
\begin{figure}[!hbt] 
\begin{center}
\includegraphics[width=0.4\textwidth,clip=]{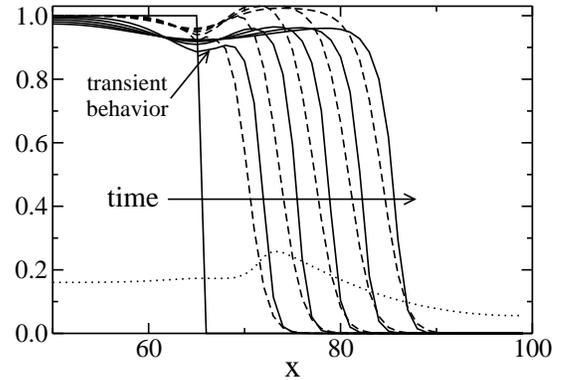} 
\end{center}
\caption{Cross-section of a tumor for the case of coupled dynamics
between matrigel $M$, tumor cells $U$ and homotype chemoattractant $C$
for different times between $t_0=\epsilon_t$ and
$t_f=30000\epsilon_t$. Solid lines: $U$ subject to homotype
chemotaxis; dashed lines: $U$ subject to no homotype
chemotaxis. Dotted line: $C$ (times a factor of four) for $t=10000
\epsilon_t$.}
\label{tumor_homo}
\end{figure}

\section{Discussion and Conclusions}
\label{sec:end}

In summary we conclude that: 

\begin{enumerate}

\item 

The matrigel $M$ induces tumor cell proliferation. This growth is an
overall expansion of the initial tumor that follows the principle of
{\em least resistance}~\cite{physicaa}. Moreover, in the case of
interest here, that of a slowly diffusing matrigel, the tumor boundary
(or surface) becomes inhomogeneous due to the random nature of the
slowly varying nutrient $M$.  It is noteworthy that the roughness of
the tumor surface depends also on the proliferation rate of the cells.

\item 

The homotype chemoattractant enhances the velocity of the tumor cells
due to the increase of $C$ at the boundary of the tumor. This effect
combined with cell proliferation (due to $M$) would induce the onset
of invasion of tumor cells towards regions of lower matrigel density.
We have been able to show that the secretion and subsequent diffusion
of $C$ {\em catalyzes} the motion of the tumor cells.

\item 

Finally, as the heterotype chemoattractant is introduced into the
system at a given distance from the tumor (in our case at the
boundary), and because it diffuses, an initially circular tumor
develops unstable invasive branches that move towards the source of
the heterotype chemoattractant. These branches develop from
{\em initial seeds} ({\em i.e.,}
fluctuations on the tumor boundary), which are due to the
mechanical confinement from $M$, and the velocity enhancement due to
$C$ and  $Q$.

\end{enumerate}

Our results are an improvement over~\cite{sander,physicaa}. We have
not limited ourselves to (i) performing a linear stability analysis
from the steady state solutions as was done in~\cite{sander} or to
(ii) numerically solving a simplified version of the
reaction-diffusion equations as carried out in~\cite{physicaa}.  On
the other hand, we have analytically and numerically studied the full
non-linear problem.  The results of the work presented here allow us
to say that proliferation is a requirement for invasion. In fact,
there can be no (chemotactic) invasion without proliferation, as
proliferation due to $M$ provides the initial seeds that trigger the
onset of invasion. That is, the slow diffusion of $M$ is crucial to
the development of those initial seeds (rough tumor interface) that
become invasive branches due to heterotype chemotactically induced
instability.

Admittedly, our model still does have several shortcomings as it
inevitably has to simplify the complex biological scenario considered
here.  For instance, tumor cell apoptosis and thus, the development of
a central necrotic core is currently not
included~\cite{necrotic}. Incorporating this characteristic tumor
feature would also have implications for the simulation
itself. Specifically, detrimental byproducts released from the dying
virtual cells would render this area ``toxic'', resulting in a central
space within the growing tumor, which is not being repopulated by the
proliferative tumor surface. In future work, this tumor characteristic
can be implemented {\em e.g.,} by some dynamic, internal boundary
condition within the tumor. We have also failed to model the finite
receptor occupancy of the tumor cell surface.  This issue is important
in order to find biological support for implementing a maximum
threshold of ({\em e.g.,} homotype) chemoattractant uptake rate (by
each tumor cell). On the other hand, the minimum threshold is given by
the maximum sensitivity of the cell surface-based receptor system.
Nonetheless, even in its present form the model already proves very
useful for interdisciplinary cancer research as it provides the
following, at least in part experimentally testable hypotheses: (i)
tumor cell proliferation by itself cannot generate the invasive
branching behaviour observed experimentally, yet, proliferation is a
requirement for invasion (ii) heterotype chemotaxis provides an
instability mechanism that leads to the onset of tumor cell invasion
and (iii) homotype chemotaxis does not provide such a mechanism but
enhances the mean speed of the tumor surface.

Combined with more specific experimental data, both on the molecular
and on the microscopic scale, this ongoing work may therefore reveal
novel and exciting insights into the role of tumor cell signaling and
its impact on the emergence of multicellular patterns.

\acknowledgments

C. \ M.-P. and T.\ S. \ D. would like to thank S. \ Habib (Los Alamos
National Laboratory) for very valuable discussions.  C.\ M.-P. would
like to thank B.\ D. \ Sleeman (University of Leeds) for a very
careful reading of the manuscript and for useful discussions.  This
work has been partially supported by MECD (Spain) Grant
No. BFM2003-07749-C05-05.  T.\ S.\ D. would like to acknowledge
support by NIH grants CA 085139 and CA 113004 and by the Harvard-MIT
(HST) Athinoula A. Martinos Center for Biomedical Imaging and the
Department of Radiology at Massachusetts General Hospital. We thank
C.\ Athale (Complex Biosystems Modeling Laboratory, Massachusetts
General Hospital) for providing the microscopy image.


\begin{thebibliography}{*99}
\bibitem{nota_eps} 

Note that the characteristic time, $\tau$, and length, $\lambda$,
scales are related to the diffusivity through the relation
$\lambda^2/\tau\sim \mu_M\sim \varepsilon^2$. This motivates a change
of variable $x\rightarrow x/\varepsilon$.
\bibitem{speed-fisher} 

Fisher's equation has a minimal speed $v_0$ and a continuum of speeds
$v > v_0$, which depend on the boundary conditions. In our case we
have a finite domain with no flux boundary conditions.
\bibitem{nota_flat}

We have placed the initial flat tumor at a constant value of the $x$
coordinate, that is: $u(x,y,t=0)=\delta(x-x_0)$. This choice and the
initial condition for $q$, Eq.~(\ref{q-init}) imply that the system of
equations only depends on $x$ and is independent of $y$. We are
dealing with an effectively one-dimensional system.
\bibitem{note_null_1}

Notice that the numerator of equation~(\ref{dcdu}) is positive (the
production term $R_C^{(p)}$ is larger than the $R_C^{(d)}$ consumption
term) and the denominator $R_M$ is always a positive function of both
arguments.
\bibitem{note_null_2}

Notice that the numerator of equation~(\ref{dcdu}) is positive but
decreases as $U$ increases (the $R_C^{(d)}$ consumption rate catches
up with the production term $R_C^{(p)}$).
\end{thebibliography}
\end{document}